\documentclass[12pt,preprint]{aastex}
\usepackage[dvips]{color}
\shorttitle{Determination of Prominence Plasma $\beta$}
\shortauthors{Hillier et al.}
\usepackage{amsmath}
\begin{document}
\title{Determination of Prominence Plasma $\beta$ from the Dynamics of Rising Plumes}

\author{Andrew Hillier\altaffilmark{1}, Richard Hillier\altaffilmark{2} and Durgesh Tripathi\altaffilmark{3}}

\email{andrew@kwasan.kyoto-u.ac.jp}

\altaffiltext{1}{Kwasan and Hida Observatories, Kyoto University}

\altaffiltext{2}{Department of Aeronautics, Imperial College}

\altaffiltext{3}{Inter-University Centre for Astronomy and Astrophysics, Post Bag 4, Ganeshkhind, Pune 411007, India }

\begin{abstract}
The launch of Hinode satellite led to the discovery of rising plumes, dark in chromospheric lines, in quiescent prominences that propagate from large ($\sim 10$\,Mm) bubbles that form at the base of the prominences.
These plumes present a very interesting opportunity to study Magnetohydrodynamic (MHD) phenomena in quiescent prominences, but obstacles still remain.
One of the biggest issues is that of the magnetic field strength, which is not easily measurable in prominences.
In this paper we present a method that may be used to determine a prominence's plasma $\beta$ when rising plumes are observed.
Using the classic fluid dynamic solution for flow around a circular cylinder with an MHD correction, the compression of the prominence material can be estimated.
This has been successfully confirmed through simulations; application to a prominence gave an estimate of the plasma $\beta$ as $\beta=0.47 \pm 0.079$ to $1.13\pm 0.080$ for the range $\gamma=1.4$\,-\,$1.7$.
Using this method it may be possible to estimate the plasma $\beta$ of observed prominences, therefore helping our understanding of a prominence's dynamics in terms of MHD phenomena.
\end{abstract}
\keywords{Magnetohydrodynamics (MHD), Sun:Prominences}

\section{Introduction}\label{intro}

Quiescent prominences are large clouds of relatively cool plasma supported against gravity by their magnetic field. 
It is known that the temperature of quiescent prominences is approximately 8000\,K \citep{TH1995} and density $\sim 10^{11}$\,cm$^{-3}$ \citep{HIR1986} which is a decrease and increase of approximately two orders of magnitude respectively from the surrounding corona. 
Using this value for the temperature the pressure scale height can be calculated as $\Lambda \sim 300$\,km, which is approximately $2$ orders of magnitude less than the characteristic height of a quiescent prominence \citep[$\sim 25$\,Mm][]{TH1995}.
There are many reviews that describe the current understanding of the nature of quiescent prominences \citep[see, for example,][]{TH1995,LAB2010,MAC2010}.

\begin{figure}[ht]
\centering
\includegraphics[height=8cm]{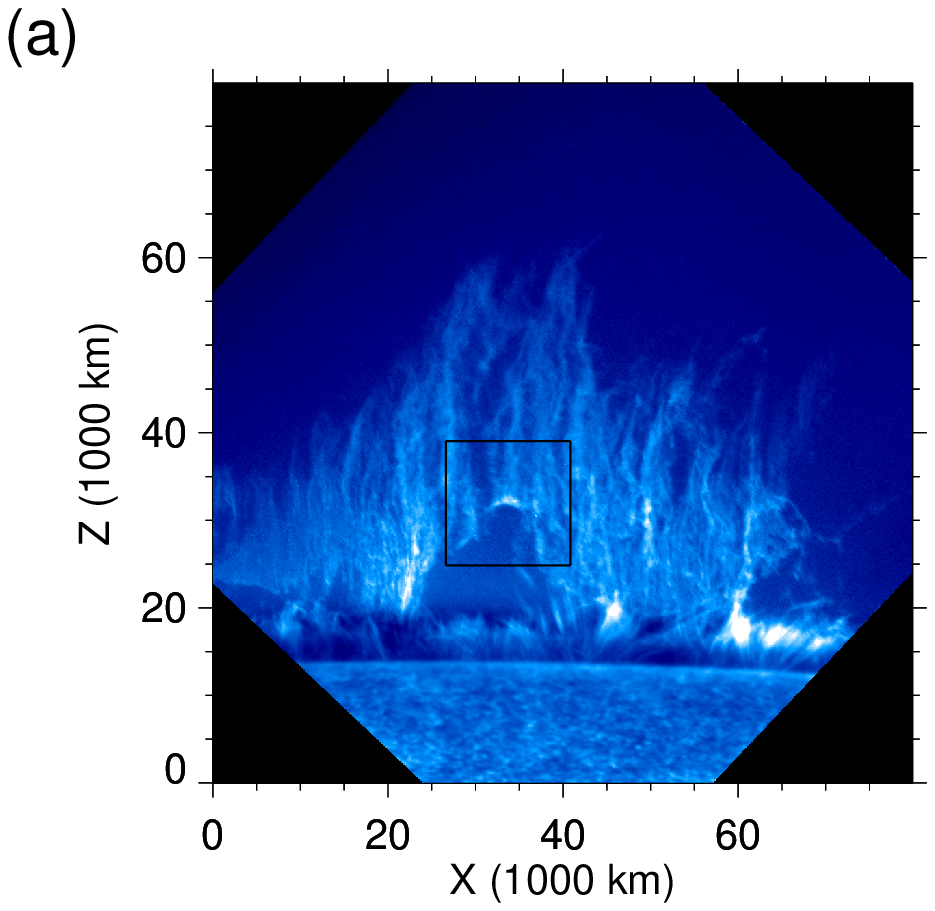}
\includegraphics[height=8cm]{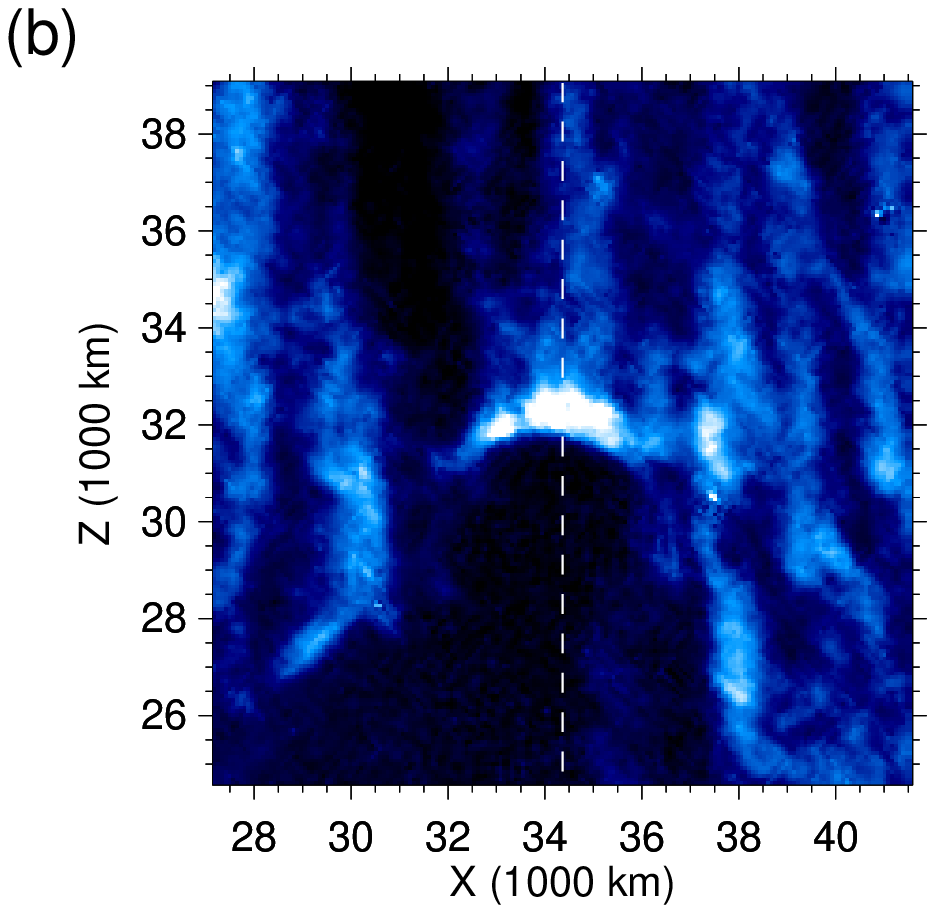}
\caption{Panel (a) shows a quiescent prominence observed on the 3-Oct-2007 at 04:16UT. Panel (b) shows a zoom-in of the region in the box in panel (a). The plume and the bright prominence plasma above it are clearly visible.}
\label{prom}
\end{figure}

Observations by the Solar Optical Telescope \citep[SOT;][]{TSU2008} on the Hinode satellite \citep{KOS2007} have shown that on very small scales quiescent prominences are highly dynamic and unstable phenomena.
\cite{BERG2008} and \cite{BERG2010} reported dark upflows that propagate from large ($\sim 10$\,Mm in size) bubbles that formed at the base of some quiescent prominences, through a height of approximately $10$\,Mm before dispersing into the background prominence material.
An example of these plumes is shown in Figure \ref{prom}.
Observations by \cite{BERG2011} show that the bubble and plumes have a minimum temperature of $250,000$\,K.
The dark upflows were found to rise at constant velocity of $\sim 20$\,km\,s$^{-1}$. 
Often these plumes would separate from the cavity forming bubbles inside the prominence material.

\cite{BERG2010} hypothesized that the observed upflows were caused by the magnetic Rayleigh-Taylor instability, as a mixed mode perturbation with an interchange mode ($k$ perpendicular to $B$) and an undular mode ($k$ parallel to $B$), in the high Atwood number limit [$A=(\rho_+ - \rho_-)/(\rho_+ + \rho_-) \rightarrow 1$], where $k$ is the wave number, $\rho_+$ is the density of the region above the contact discontinuity and $\rho_-$ is the density of the below the contact discontinuity.
\cite{RYU2010} described how the theoretically predicted growth rates and behaviour for the magnetic Rayleigh-Taylor instability well match the observations of quiescent prominence plumes.
\cite{HILL2011} and \cite{HILL2012} showed how upflows could be created by the 3D mode of the magnetic Rayleigh-Taylor instability acting on the boundary between the Kippenhahn-Schl\"{u}ter prominence model \citep{KS1957} and a hot tube inserted into the it.
The upflows created were found to be qualitatively similar to the observed plumes.

Observations of the magnetic field of quiescent prominences show the field strength to be in the range $\sim 3$\,-\,$30$ G \citep{LER1989}, where the average for polar-crown prominences is $\sim 5$\,G \citep{ANHEIN2007}.
However, as it is very difficult to observe the prominence magnetic field directly, often other methods to determine the magnetic field of prominences have been employed.
One common mechanism has been to extrapolate the photospheric magnetic field to obtain values for the coronal field where the prominence is observed \citep{DUD2008,AUL2003}.
Another common method is to analyse the oscillatory pattern of filaments \citep{ISO2006, PINT2008, TRIP2009} or filament threads \citep{LIN2007, LIN2009} to calculate the magnetic field through the application of wave theory.
This is known as prominence seismology \citep{AGG2012}. 
In this paper we present a new method to investigate the magnetic field strength relative to the gas pressure by determining the plasma $\beta$ (ratio of gas pressure to magnetic pressure) of a prominence through the study of compression of the prominence material by the Rayleigh-Taylor plumes.

\section{The Observations Needed to Determine a Prominence's Plasma $\beta$}

Before the methodology through which the prominence plasma $\beta$ can be determined is presented, first we will quickly review the observational characteristics of the prominence plumes that are important for this analysis.
The points we need to remember are \citep{BERG2010}:
\begin{itemize}
\item The plumes have a phase in their evolution where they rise at an approximately constant velocity

\item The plumes have an approximately circular (elliptical) head

\item The intensity of the prominence material at the top of the plume is greater than the average prominence intensity.

\item Flows of material can be observed along the head of the plume

\end{itemize}

To make it possible to use these observables, a number of assumptions are necessary.
The first is in relation to the density of the prominence.
The prominence intensity is known to be a result of scattering.
Therefore the intensity is directly related to the column density of the prominence.
This allows us to assume that the intensity can be used as a proxy for the density.
Hence the bright top of the plume would imply that there has been an increase in the density through compression of the prominence.

The next assumption is in relation to the modelling of the geometry of the problem.
As the plume moves at an approximately constant velocity with an elliptical head, this can be used to simplify the system under study.
The first step is to remove the stem of the plume, and just leave an ellipse in the prominence material.
Then as this ellipse is moving at constant velocity through (what we assume to be) a constant medium, a shift in reference frame can make the situation of an ellipse, stationary in a constant flow.
The effect of a non-stationary prominence will be investigated to some extent later.
Now we must consider the 3D nature of the geometry under investigation.
The observed plumes are known to be caused by the magnetic Rayleigh-Taylor instability \citep{BERG2010,HILL2012}
It is also known that the 3D magnetic Rayleigh-Taylor instability creates filamentary structures aligned with the magnetic field \citep{ ISO2005, ISO2006b, Stone2007,HILL2012} or the bi-sector of the upper an lower magnetic fields in the case where shear is present \citep{Stone2007}.
Therefore we can view the problem as being similar to a elliptical cylinder in a constant flow.
Then a coordinate transform to cylindrical coordinates would effectively give a circle in a constant flow.
This has now reduced to a problem that has been significantly studied in fluid dynamics.

The transform described above is a conformal transform of the coordinate system.
In an incompressible regime, such a conformal transform would automatically give the correct result, but in a compressible regime, where nonlinearity is important, the velocity will not be completely accurate.
The smaller the eccentricity and the fast-mode Mach number, the more accurate the coordinate transform will be.

\section{Derivation of equations to determine a prominence's plasma $\beta$}

Here we use a classic result from fluid dynamics to define a set of equations that can be used to interpret prominence observations and determine physical parameters relating to that prominence.
First, we describe the thought process behind the method, followed by a derivation of the equations used to determine the plasma $\beta$ of the prominence.

%\subsection{Formulation of velocity field for a potential flow around a circular cylinder in a magnetized plasma}

The derivation presented here is analogous to that presented in \citet{VD1975} for a compressible flow around a circular cylinder in a steady state.
Here the derivation is altered to include a magnetic field that is perpendicular to the plane in which the plasma is flowing.
In this derivation we assume that the magnetic field is in the $x$\,-\,direction and the flow has velocity components in the $y$\,- and $z$\,-\,direction.
All derivatives in the $x$\,-\,direction are assumed to be $0$.
The validity of these assumptions will be investigated through comparisons with simulations in section \ref{SIM}.
This analysis uses the equations of mass conservation, momentum and energy.

The equation of mass conservation, for a steady planar two-dimensional flow is:
\begin{equation}
\frac{\partial \rho v}{\partial y}+\frac{\partial \rho w}{\partial z}=0
\end{equation}
where $\rho$, $v$ and $w$ are the density, the $y$\,-\,component of the velocity and the $z$\,-\,component of the velocity.
The equation presented above is for a steady state solution, so there is no temporal variation of the density and the velocity field.
All these values have been normalised by the free stream reference values for density $\rho_{\infty}$ and velocity $U_{\infty}$, where the values with the subscript $\infty$ denote the values upstream of the plume head.
This can then be rewritten to give:
\begin{equation}\label{mass_cons}
\frac{\partial v}{\partial y}+\frac{\partial w}{\partial z}=-\frac{v}{\rho}\frac{\partial \rho}{\partial y}-\frac{w}{\rho}\frac{\partial \rho}{\partial z}.
\end{equation}

If the flow is irrotational, then the momentum equations can be simply replaced by the statement that the vorticity is zero, that is:
\begin{equation}\label{irrot}
\frac{\partial w}{\partial y}-\frac{\partial v}{\partial z}=0
\end{equation}
which is satisfied by the velocity potential $\phi$ where:
\begin{equation}
v=\frac{\partial \phi}{\partial y}; w=\frac{\partial \phi}{\partial z}
\end{equation}

The next step is to relate $\rho$ to the velocity, so that Equation \ref{mass_cons} can be expressed purely in velocity derivatives, that is to say purely in derivatives of $\phi$.
For a perfect gas under adiabatic contraction/expansion that has a magnetic field perpendicular to the plane of the flow, the MHD-Bernoulli's equation can be written as:
\begin{equation}
\frac{1}{2}\left(v^2+w^2\right)+\frac{1}{\gamma-1}\frac{p\gamma}{\rho}+\frac{B_x^2}{4\pi\rho}=C(S)
\end{equation}
where $C(S)$ is used to show that the value is constant along streamlines.
Therefore, the values of the physical variables at each point in the flow can related to the free stream values (the values of the flow at an infinite distance from the circular cylinder).
\begin{equation}\label{ENERGY}
\frac{\gamma}{\gamma-1}\frac{p p_{\infty}}{\rho \rho_{\infty}}+\frac{1}{2}\left( v^2 +w^2 \right)U^2_{\infty}+\frac{B_x^2}{\rho}\frac{B_{x \infty}^2}{4\pi \rho_{\infty}}=\frac{\gamma}{\gamma-1}\frac{p_{\infty}}{\rho_{\infty}}+\frac{1}{2}U^2_{\infty}+\frac{B_{x \infty}^2}{4\pi \rho_{\infty}}
\end{equation}
where the subscript $\infty$ implies the free stream values that are used for normalisation.

If we define the free stream speed of sound as $C_s=\sqrt{\gamma p_{\infty}/\rho_{\infty}}$, the free stream Alfven velocity as $V_a=B_{x\infty}/\sqrt{4\pi \rho_{\infty}}$ and denote the free stream Mach number as $M_{\infty}=U_{\infty}/C_s$ then equation \ref{ENERGY} becomes:
\begin{equation}
\frac{1}{\gamma-1}\frac{p}{\rho}+\frac{1}{2}M_{\infty}^2(v^2+w^2)+\frac{B_x^2}{\rho}\frac{V_A^2}{C_s^2}=\frac{1}{\gamma-1}+\frac{1}{2}M_{\infty}^2+\frac{V_A^2}{C_s^2}.
\end{equation}
We can then define a relation between the Alfven velocity and the sound speed as follows:
\begin{equation}
\frac{V_A^2}{C_s^2}=\frac{2}{\gamma \beta_{\infty}}
\end{equation}
where $\beta_{\infty}=B_{x\infty}^2/p_{\infty}$ is the free stream plasma $\beta$.
This then gives:
\begin{equation}\label{Mach_eqn}
\frac{1}{\gamma-1}\frac{p}{\rho}+\frac{1}{2}M_{\infty}^2(v^2+w^2)+\frac{B_x^2}{\rho}\frac{2}{\gamma \beta_{\infty}}=\frac{1}{\gamma-1}+\frac{1}{2}M_{\infty}^2+\frac{2}{\gamma \beta_{\infty}}.
\end{equation}

As we have assumed that the flow is irrotational, we can assume that the flow is constant entropy.
For a perfect gas (in normalised form) this can be expressed as: $p/\rho^{\gamma}=1$.
We can also use the identity that $B_x^2=p/\beta'$, where $\beta' \beta_{\infty}$ defines the plasma $\beta$ of system.
Using these identities, equation \ref{Mach_eqn} becomes
\begin{equation}
\rho^{\gamma-1}+\frac{\gamma-1}{2}M_{\infty}^2(v^2+w^2 -1)+\frac{\rho^{\gamma-1}}{\beta'}\frac{2(\gamma-1)}{\gamma \beta_{\infty}}=1+\frac{2(\gamma-1)}{\gamma \beta_{\infty}}
\end{equation}
To tidy up the equations, we define:
\begin{eqnarray}
D(y,z)=1+\frac{1}{\beta'}\frac{2(\gamma-1)}{\gamma \beta_{\infty}}\\
C=1+\frac{2(\gamma-1)}{\gamma \beta_{\infty}}
\end{eqnarray}
giving:
\begin{equation}
\rho^{\gamma-1}+\frac{\gamma-1}{2}\frac{1}{D(y,z)}M_{\infty}^2(v^2+w^2 -1)=\frac{C}{D(y,z)}
\end{equation}

For simplicity, we will now make the assumption that the plasma $\beta$ is approximately uniform.
In reality, we would have to solve for this using the induction equation, but for simplicity this is neglected in this analysis.
This leads to the following equation:
\begin{equation}\label{reduced_en}
\rho^{\gamma-1}+\frac{\gamma-1}{2}\frac{1}{D}M_{\infty}^2(v^2+w^2 -1)=1.
\end{equation}
Now it is possible to combine equation \ref{reduced_en} with equation \ref{mass_cons} to determine an equation for the velocity field.

If we follow the derivation by \citet{VD1975}, by transforming to cylindrical coordinates (where the length-scale is normalised by the radius of the cylinder) we can then calculate the velocity potential $\phi$ (with all $\partial / \partial x=0$):
\begin{equation}\label{firstMstar}
\phi=\left(r+\frac{1}{r} \right)\cos \theta - {M_*}^2\left[ \left(\frac{13}{12}\frac{1}{r}-\frac{1}{2}\frac{1}{r^3}+\frac{1}{12}\frac{1}{r^5} \right) \cos\theta -\left(\frac{1}{12}\frac{1}{r^3} - \frac{1}{4}\frac{1}{r} \right)\cos ^3 \theta   \right]
\end{equation}
where:
\begin{equation}\label{Meqn}
M_* = \sqrt{\frac{1}{D}}M_{\infty}
\end{equation}
which can be viewed as the fast-mode Mach number.
Using $\mathbf{v}= \nabla \phi$, we get:
\begin{align}
v_r(r,\theta) = & \left(1-\frac{1}{r^2} \right)\cos \theta - {M_*}^2\left[ \left(-\frac{13}{12}\frac{1}{r^2}+\frac{3}{2}\frac{1}{r^4}-\frac{5}{12}\frac{1}{r^6} \right) \cos\theta +\left(\frac{1}{4}\frac{1}{r^4} - \frac{1}{4}\frac{1}{r^2} \right)\cos ^3 \theta   \right] \notag \\
& + O({M_*}^4)\label{vr}\\
v_{\theta}(r,\theta) = & -\left(1+\frac{1}{r^2} \right)\sin \theta + {M_*}^2\left[ \left(\frac{13}{12}\frac{1}{r^2}-\frac{1}{2}\frac{1}{r^4}+\frac{1}{12}\frac{1}{r^6} \right) \sin\theta -\left(\frac{1}{12}\frac{1}{r^4} - \frac{1}{4}\frac{1}{r^2} \right)3 \cos ^2 \theta \sin \theta   \right] \notag \\
& + O({M_*}^4)\label{vthe}
\end{align}

\subsection{Derivation of density distribution}

It is now necessary to derive an equation for the distribution of the density.
We need to convert equation \ref{reduced_en} to cylindrical coordinates:
\begin{equation}\label{derivedeqn}
\rho^{\gamma-1}+\frac{\gamma-1}{2}{M_*}^2(v_r^2+v_{\theta}^2 -1)=1
\end{equation}
Rewriting Equation \ref{derivedeqn}, we find:
\begin{equation}\label{dendist}
\rho=\left(1-\frac{\gamma-1}{2}{M_*}^2(v_r^2+v_{\theta}^2 -1)\right)^{1/(\gamma-1)}
\end{equation}
This equation can now be used to determine the expected density distribution around a plume head.

\begin{figure}[ht]
\centering
\includegraphics[height=8cm]{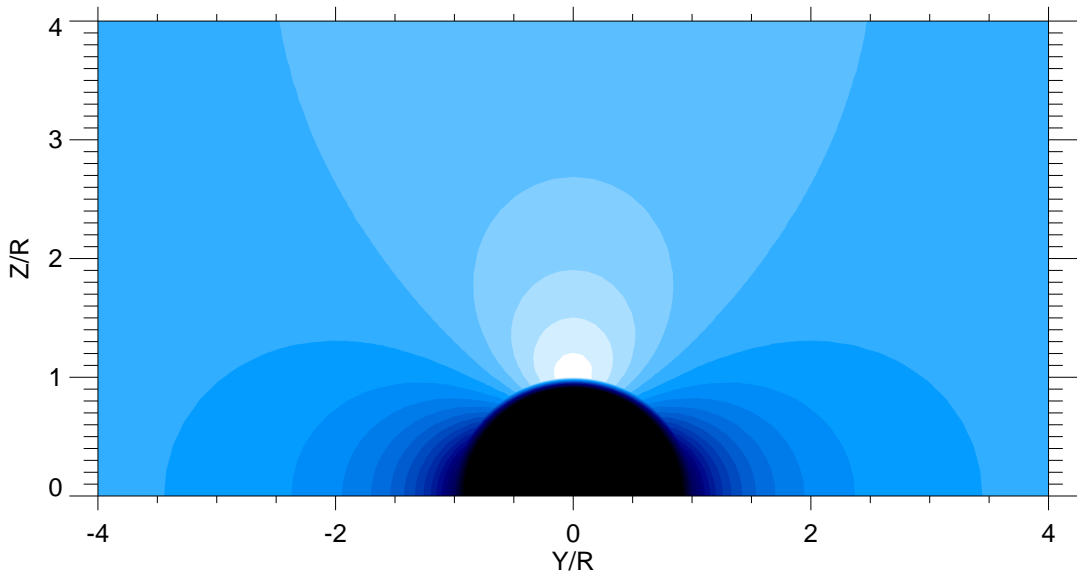}
\caption{The density found for a compressible flow around a circular cylinder. Brighter colours imply higher densities.}
\label{eqnden}
\end{figure}

Figure \ref{eqnden} shows the density distribution around a circular cylinder.
The high density region, created as a result of the compression of the plasma, is clearly visible at the top of the circular cylinder.
This is for the case where the higher order terms were used, i.e. the distribution shown is from Equation \ref{dendist}.

\section{Comparison with Simulation Results}\label{SIM}

As it is not possible to directly confirm the effectiveness of this method for determining the plasma $\beta$ of a prominence, it is important to apply it to simulations results to provide some validation.
We apply it to the results of \citet{HILL2011} and \citet{HILL2012}, where a full description of the simulations are presented.
In this section we apply Equation \ref{dendist} to solve the forward problem, i.e. we use the parameters of the simulation - including the plasma $\beta$ - to see how well the density distribution around the plume head can be reproduced.

%\subsection{The model used for the simulation}

\begin{figure}[ht]
\centering
\includegraphics[height=8cm]{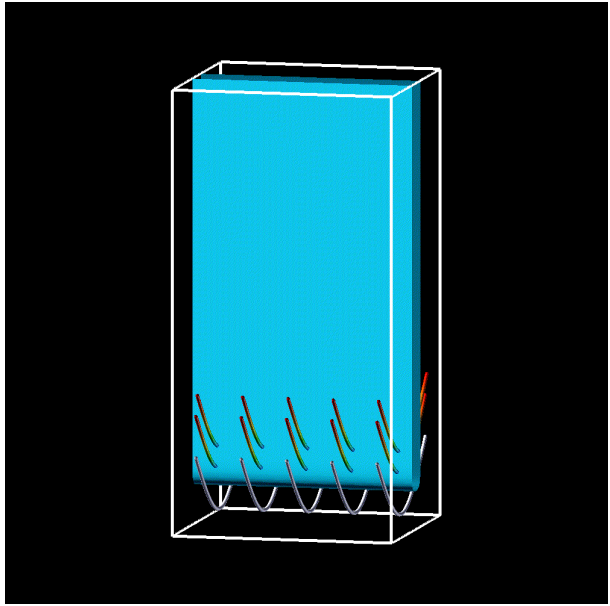}
\caption{3D rendering of the initial conditions of the simulation. The blue isosurface shows the density and the lines show the magnetic field.}
\label{INIT}
\end{figure}

The simulation used in this paper is of the 3D magnetic Rayleigh-Taylor instability in the Kippenhahn-Sch\"{u}lter prominence model.
To destabilise the Kippenhahn-Sch\"{u}lter prominence model, a low density tube (density of 30\% of the prominence) was placed inside the model.
The boundary between this tube and the dense material was subjected to a random velocity perturbation allowing the magnetic Rayleigh-Taylor instability to develop.
A 3D rendering of the initial conditions of this simulation are shown in Figure \ref{INIT}.

In this study, 3D conservative ideal MHD equations were used. 
Constant gravitational acceleration was assumed, but viscosity, diffusion, heat conduction and radiative cooling terms were neglected and an ideal gas was assumed.
The equations were non-dimensionalised using the sound speed ($C_s=13.2$\,km\,s$^{-1}$), the pressure scale height ($\Lambda=C_s/(\gamma g)= R_gT/(\mu g)=6.1 \times 10^7$ cm), the density at the centre of the prominence ($\rho (x=0)=10^{-12}$ g cm$^{-3})$ and the temperature ($T_0=10^4$\,K), giving a characteristic timescale of $\tau=\Lambda/C_s=47$\,s.
The ratio of specific heats was set as $\gamma=1.05$ and $\beta=0.5$.
A full description of the simulations can be found in \citet{HILL2012}.

\subsection{Description of simulated plumes}\label{SIMPlume}

\begin{figure}[ht]
\centering
\includegraphics[height=8cm]{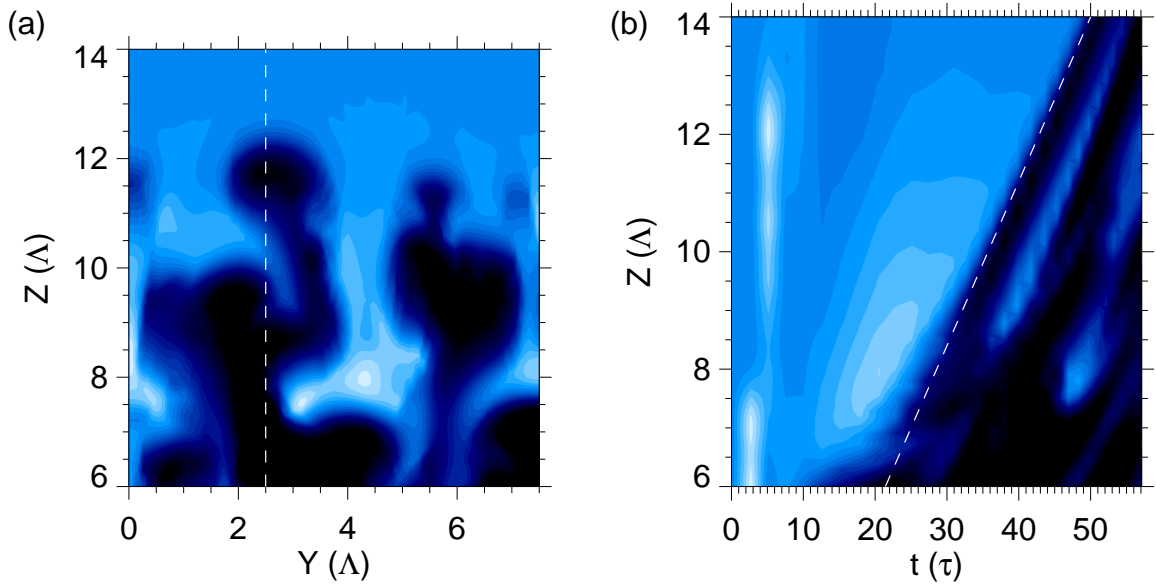}
\caption{Panel (a) shows the rising plumes that are formed by the magnetic Rayleigh-Taylor instability in the Kippenhahn-Schl\"{u}ter model. Panel (b) shows the time-distance diagram taken along the dashed line in panel (a). This allows the Mach number of the rise velocity of the plume to be calculated as $M=0.28$.}
\label{plumerise}
\end{figure}

Panel (a) of Figure \ref{plumerise} shows the density distribution at the centre of the the Kippenhahn-Schl\"{u}ter model as the rising plumes, created by the magnetic Rayleigh-Taylor instability move upward through the dense prominence material.
The plume in the box, more like a bubble in this image but really a low-density tube, is the plume we use for this comparison.
Note that we didn't have to choose a plume that has detached from its tail, as we are only interested in the density at the head of the plume.

Panel (b) of Figure \ref{plumerise} shows the time-distance plot taken along the slit in Panel (a).
The constant rise velocity of the plumes, resulting from the force balance created at the head of the plume, is clearly shown \cite{HILL2012}.
The rise velocity of the topmost plume was found to have a Mach number of $M=0.28$.
We use the free-stream Mach number because if we transform the plume head to 0-velocity rest frame then the rise velocity transforms to a free-stream velocity of the dense material.

\begin{figure}[ht]
\centering
\includegraphics[height=8cm]{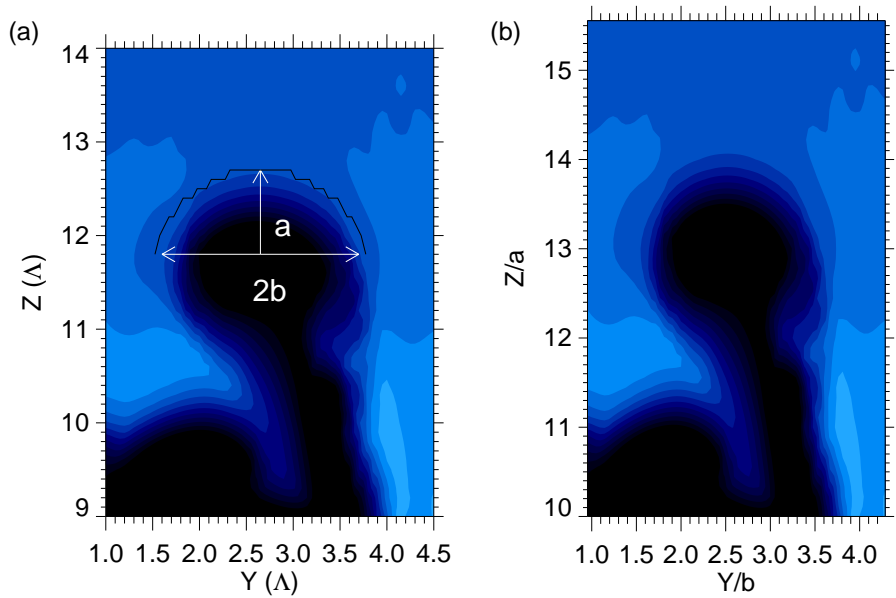}
\caption{Calculation of the coordinate shift for the plume head. panel (a) shows the original size with the plume dimensions marked ($a=1.05\Lambda$ and $b=0.9\Lambda$) and panel (b) shows the result of that transform. The arch at the top of the plume marks the size of the plume head.}
\label{transform}
\end{figure}

It can be instantly noticed in Figure \ref{plumerise} that the plume is not perfectly circular.
The first step of the transform is to change the coordinate size so that the size of the elliptical bubble is that of a unit ellipse (i.e. the area of the ellipse is $\pi$).
This is achieved by dividing the $y$ and $z$ axis by $\sqrt{ab}$, where $a$ is the minor axis of the ellipse and $b$ is the major axis of the ellipse.
A second transform needs to be performed to transform the unit ellipse to a unit circle.
This is achieved by multiplying the $z$ axis by $\sqrt{b/a}$ and the $y$ axis by $\sqrt{a/b}$.
This process is shown in Figure \ref{transform}.
The curved line marking the head of the plume is used in Figure \ref{plumeheadvel} to compare the velocity distribution at the plume head to the expected distribution.

\subsection{Comparison of Simulated and Predicted Density Distribution and Velocity Field}\label{eqnapp}

To apply Equation \ref{dendist}, various parameters have to be determined.
The Mach number of the rising flow is shown in Figure \ref{plumerise} to be $0.28$.
The density distribution needs to take in mass conservation, so the column density (integrating in the x-direction) needs to be used.
We calculate the average density over $x=0$\,-\,$2.5\Lambda$, where x is the direction of the horizontal magnetic field in the simulation domain.
In a similar method to determining the density distribution, the average plasma $\beta$ is calculated in the same way, giving $\beta \sim 0.55$.
The ratio of specific heats ($\gamma$) was taken as $\gamma=1.05$ for the simulations presented here.

\begin{figure}[ht]
\centering
\includegraphics[height=8cm]{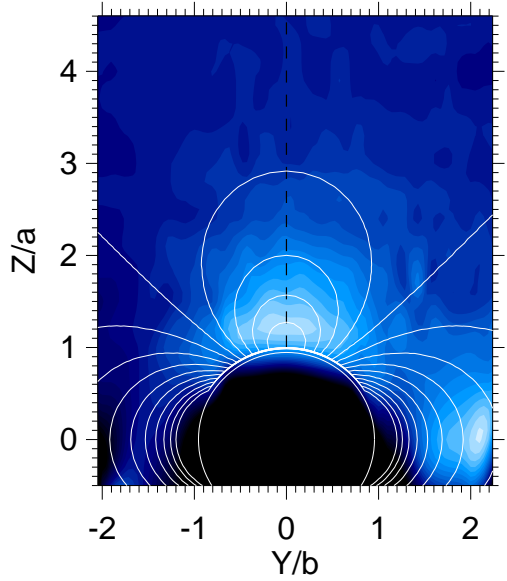}
\caption{This shows the average density (averaging in the x-direction). The density increase at the head of the plume can be clearly seen. The white contours show the predicted distribution. The dashed black line shows the position of the slit used in Figure \ref{denprofile}. The background distribution has been removed.}
\label{avden}
\end{figure}

The next important step is to determine the background values.
We determine the background density profile as the density profile at the same height at the dashed line along the left-hand edge of Figure \ref{avden}.
This is then removed from the distribution above the head of the plume to show the density increase due to the compression by the rising plume.
Finally the background distribution, required to normalise the density distribution, is calculated from the average distribution (over $x\in [0,2.5\Lambda]$) as $\rho_{av}\sim0.67$.
One final point to note is that the background distribution has a weak upflow (Mach number $M=0.07$) that needs to be included in the calculation to move the rising plume into a stationary reference frame with a free stream Mach number of $M_{\infty}=0.21$.
It is important to note that because the frame of reference is changed, this free stream Mach number is considered as the Mach number of the flow toward a stationary plume head.

Figure \ref{avden} shows the density profile of the simulation at the head of the plume, with contours from the predicted distribution to highlight the expected distribution.
The high density region at the head of the plume and the rarefied regions at the sides of match well between the predicted and simulated densities.

\begin{figure}[ht]
\centering
\includegraphics[height=8cm]{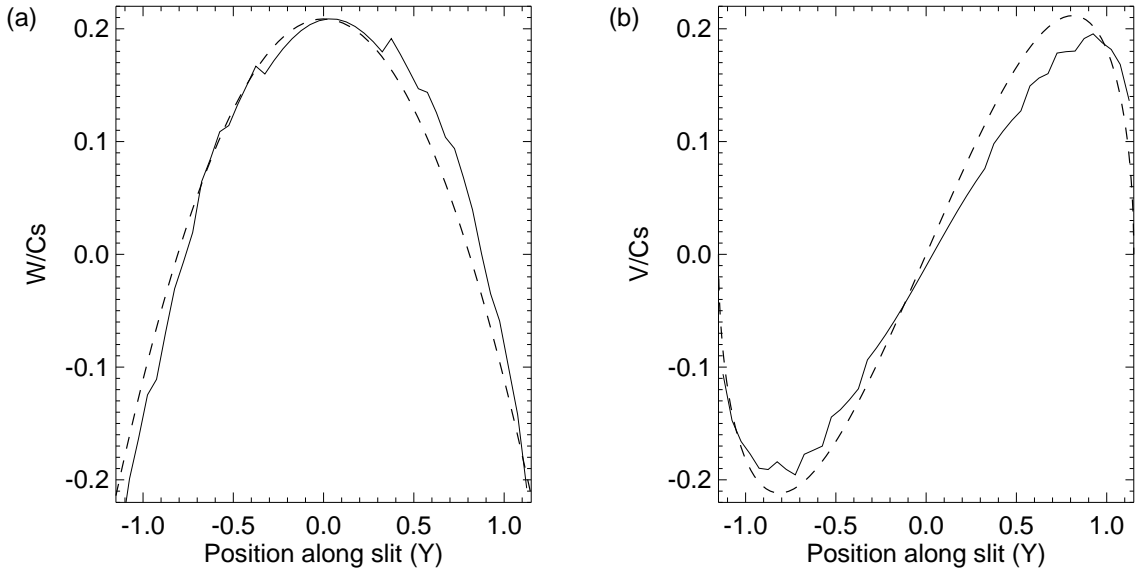}
\caption{Velocity distribution along the head of the plume (as marked in Figure \ref{transform} panel (a)).The solid line shows the simulated distribution and the dashed line shows the predicted distribution.}
\label{plumeheadvel}
\end{figure}

Next we compare the velocity distribution at the plume head with the distribution predicted by Equations \ref{vr} and \ref{vthe}.
Figure \ref{plumeheadvel} shows the predicted distributions (solid line) and the simulated distributions (dashed line) for the vertical and horizontal velocity.
The horizontal scale uses the same scale as that of Figure \ref{transform}, but centers the zero position at the plume head.
The velocity distributions are in general good agreement.
It should be noted that the velocities are multiplied by a factor of $\rho(x=0)/\rho_{av}$ to give the correct magnitude of the velocity (i.e. the maximum value of $W/Cs \sim 0.21$ ).
Note the good agreement in the velocity field even though the circular solution has been modified to apply to an elliptical body. 

\begin{figure}[ht]
\centering
\includegraphics[height=8cm]{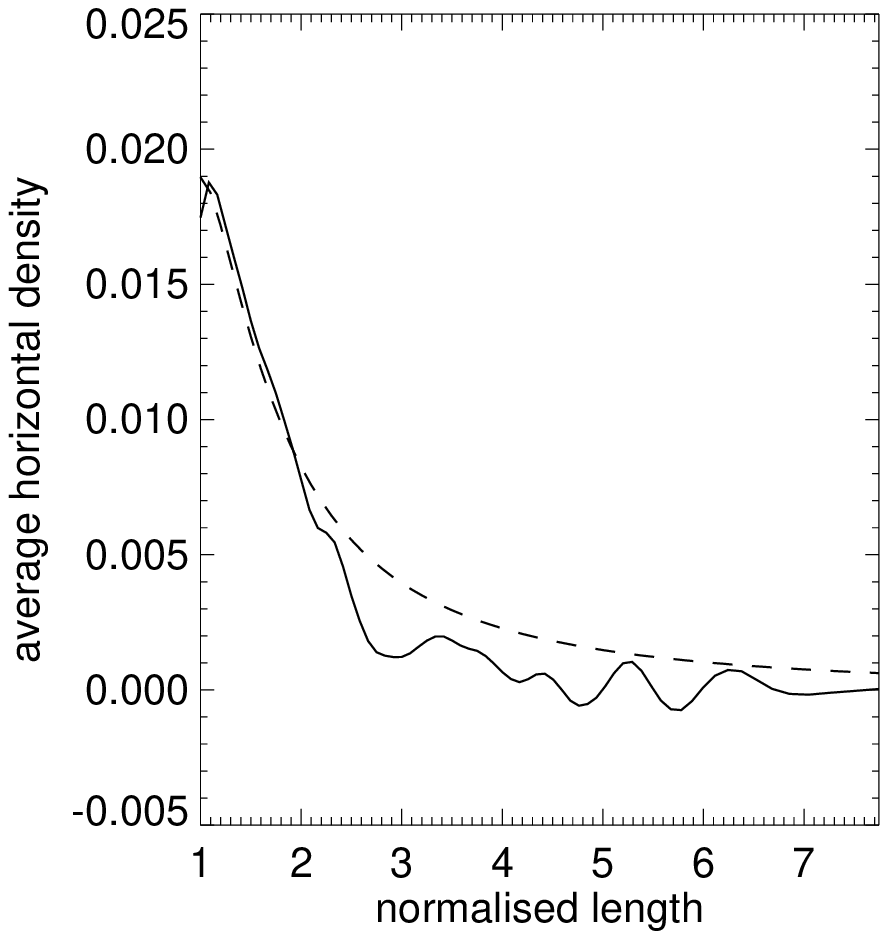}
\caption{Density profile along dotted line shown in Figure \ref{avden} minus the background density distribution (shown with the solid line) and the expected distribution using the parameters of the simulation explained in Section \ref{eqnapp} (shown with dashed line).}
\label{denprofile}
\end{figure}

Figure \ref{denprofile} shows the density distribution (minus the background distribution) at the top of a plume (solid line).
The dashed line shows the expected density distribution when using the parameters discussed above to give a value of $M_* \approx 0.19$.
The plots demonstrate that the theoretically predicted density distribution and velocity field well describes those from the simulation.

\begin{figure}[ht]
\centering
\includegraphics[height=8cm]{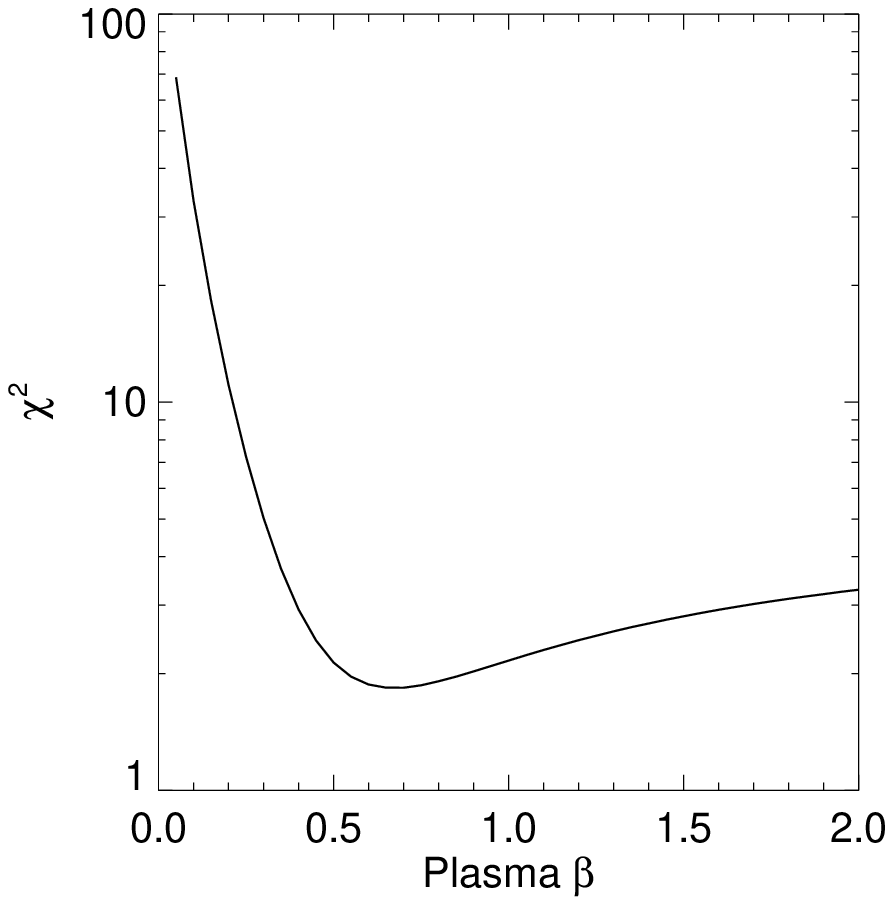}
\caption{$\chi^2$ distribution calculate for different values of plasma $\beta$ giving different density distributions and the simulated distribution.}
\label{chisqu}
\end{figure}

To investigate this further, we can fit the density distribution using Equation \ref{dendist} and solve for ${M_*}^2$.
This, in turn, will make it possible to determine the value of the plasma $\beta$.
The method is explained in greater detail in the next section.
The fit for the density distribution gives ${M_*}^2=0.038$, which gives a plasma $\beta$ of $\beta = 0.59$ when solving for the unknown (i.e. $\beta$) in Equation \ref{Meqn}.
This value is slightly higher than the input value, but sufficiently close to show that the method can estimate the plasma $\beta$ using the compression of the plume.
Figure \ref{chisqu} shows the chi-squared for the density distribution for models using different plasma $\beta$ values and the simulated distribution.
The minimum of the $\chi^2$ distribution at a $\beta$ value that is close the value of plasma $\beta$ in the simulation.

\section{Application to Observed Prominence Data}\label{PROM_SEIS}

In this section we apply the method presented above to determine the plasma $\beta$ of the plume displayed in Figure \ref{prom}.
In contrast to Section \ref{SIM}, this is treated as an inverse problem, where using the intensity and rise velocity allows the plasma $\beta$ of the prominence to be deduced.
Due to the timescales of these compressions ($\sim100$\,s) being shorter than the timescales for the cooling ($\sim 1000$\,s), we can assume to first-order approximation that the increased intensity comes through purely adiabatic effects.
These adiabatic effects would be an increased column density and a temperature rise from the compression.
Both of these can be calculated from the model described above.

Figure \ref{prom} panel (a) shows a quiescent prominence seen on the NW solar limb on 2007 October 03 observed by the SOT with the Ca {\footnotesize II} H filter at a cadence of 30 s.
The time series of this observation was between 01:16UT and 04:59UT.
This prominence presents many interesting dynamic features, for example the start of this observation (01:16UT) a large bubble has formed inside the prominence similar to those described in \cite{BERG2010}.
There are also a number of bright threads and downwardly propagating knots that occur during the duration of the observations as well as the upwardly ejected plasma blobs presented in \cite{HILL2011}.
In this paper we are interested in the rising plume and, in particular, the bright rim that forms at the top of the plume.
This is shown in panel (b) of Figure \ref{prom}. 

\subsection{Calculation of the Plume Mach Number}\label{MACH}

Before the plasma $\beta$ is calculated, it is necessary to calculate the Mach number of the rising plume.
Following the method applied in Section \ref{SIMPlume}, we create a time-distance plot to show how the height of the plume changes with time.
The dashed vertical line in Panel (a) of Figure \ref{prom} shows the position of the slit used.

\begin{figure}[ht]
\centering
\includegraphics[height=8cm]{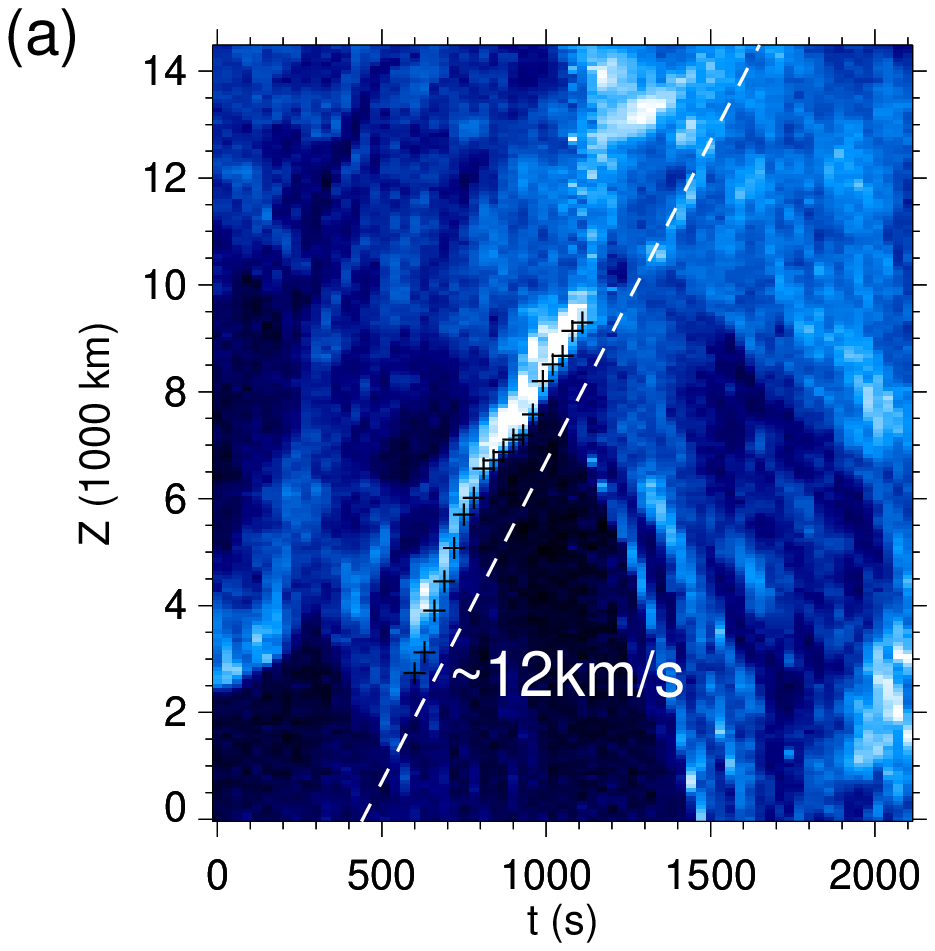}
\includegraphics[height=8cm]{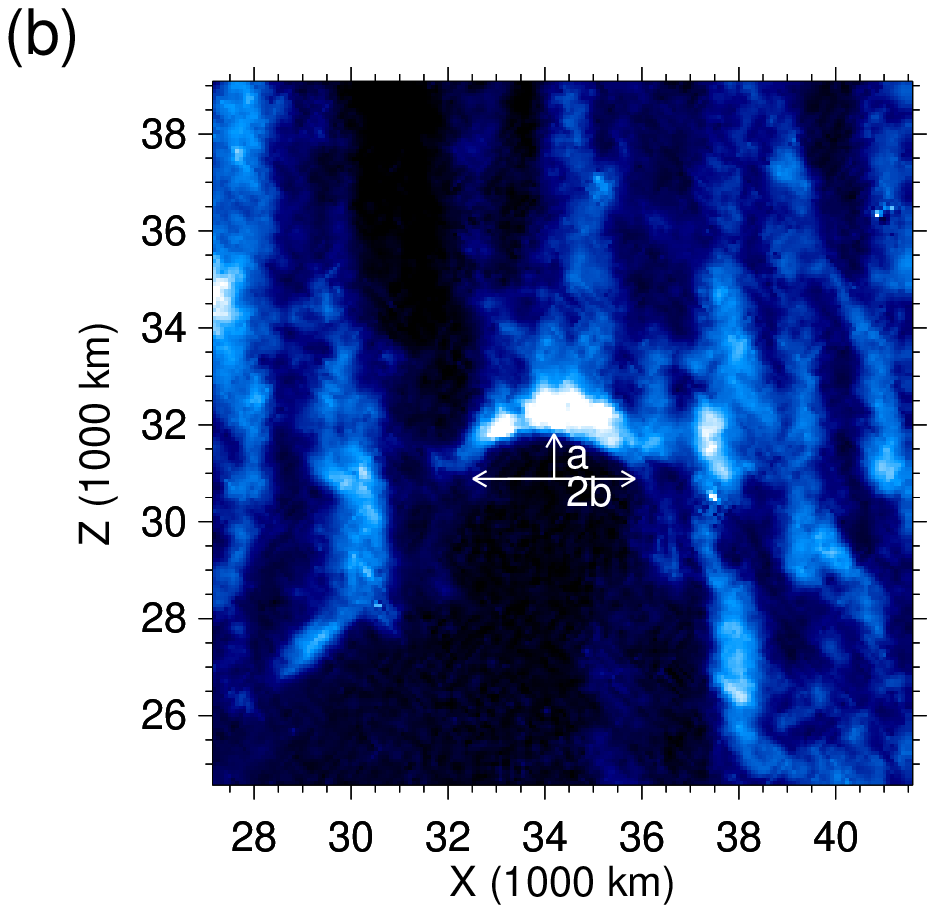}
\caption{Panel (a) shows time-distance diagram along the slit shown in Figure \ref{prom} panel (b). The cross marks show the position of the top of the plume. The rise velocity of the plume is found to be $\sim 12$\,km\,s$^{-1}$. Panel (b) shows the dimensions of the ellipse used to model the rising plume.}
\label{timedist}
\end{figure}

Panel (a) of Figure \ref{timedist} shows the time-distance plot, where the bright, rising region is the rising lip of the plume.
The cross marks show the top of the plume.
This is calculated as the pixel with the greatest intensity change from the pixel below change multiplied by the pixel intensity.
Using these pixels, the velocity of the plume and the velocity error were calculated to be $v_{plume}=12.3 \pm 0.6$\,km\,s$^{-1}$.
As this rise velocity is calculated from a time-distance diagram, it should be viewed as a lower limit on the velocity.
The dashed line in the figure shows a velocity of $\sim 12$\,km\,s$^{-1}$.
Throughout the rise of the plume, the rise velocity can be seen to be approximately constant.

To calculate the Mach number of a flow it is necessary to know the sound speed of the ambient material.
The sound speed of the dense prominence material is $C_s=[\gamma (R/\mu)T_{prom}]^{1/2}=11$\,km\,s$^{-1}$ for $\gamma=5/3$, $R=8.3\times 10^7$\,erg\,K$^{-1}$\,mol$^{-1}$, $\mu=0.9$ and $T_{prom}=8000$\,K.
Using the velocity calculated above we find the Mach number of the flow to be $M_{\infty}=1.12 \pm 0.05$ in the case where $\gamma=5/3$.

Panel (b) of Figure \ref{timedist} shows the dimensions of the ellipse used.
In this case $a=1.3$\,arcsec and $b=2.35$\,arcsec.
These lengthscales can now be used to determine the normalising length scales for the system, so that the plume head becomes circular.
Now the prominence plasma $\beta$ can be determined.

\subsection{Determining Plasma $\beta$ of Observed Prominences}

\begin{figure}[ht]
\centering
\includegraphics[height=8cm]{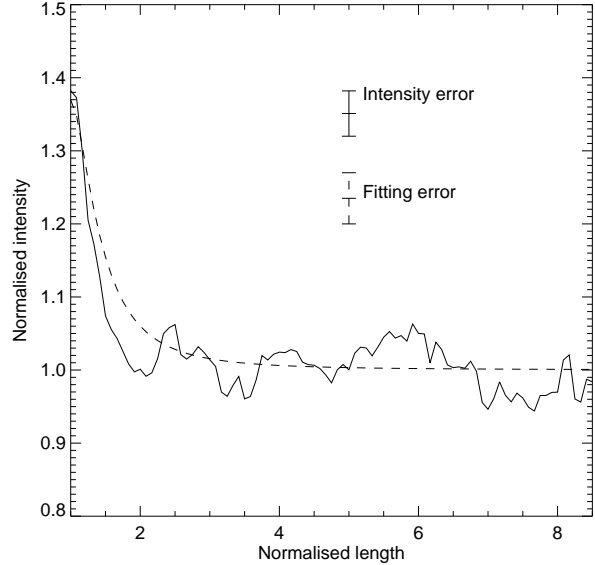}
\caption{The graph shows a plot of intensity along the slit (solid line) and the fitting assuming the intensity proportional to density (dashed line) for $\gamma=1.65$. Error bars equivalent to twice the standard deviation are presented for the observed intensity and the fit.}
\label{intfit}
\end{figure}

Figure \ref{intfit} shows the normalised intensity.
Using the emissions for heights greater than $4$ in Figure \ref{intfit} a linear fit was used to de-trend the data (as was applied to the simulations).
This intensity is used as a proxy for the column density assuming the emission is scattering dominate ($\propto \rho$).
In this paper we do not take into account the change in emission from the change in temperature.
The fitting for the density is shown by the dotted line.
This fit is performed using the IDL routine curvefit.pro.
The fit solves for the value of ${M_*}^2$ giving the values shown in Table \ref{gam_table}.

It is important to determine the error associated with this fit, so that the error in the plasma $\beta$ can later be calculated.
There are two important errors that are associated with this fit: the intensity error and the error between the observed curve and the model curve.
These errors then determine the error for ${M_*}^2$.

The intensity error is determined in a simple fashion.
The fluctuations of intensity in the corona above the prominence are used.
Here we assume that the fluctuations in the Ca II H line (which should not have a coronal signal) are due to stray light, and so we assume this is representative of the error in the intensity.
For a region in the corona above the prominence of size [200,100] pixel taken over 100 consecutive frames, the standard deviation of the intensity fluctuations is given as $\sigma_{cor}=1.33$.
As the fitting is normalised, the error should also be normalised.
Performing this normalisation results in a normalised standard deviation of $\hat{\sigma}_{cor}=0.028$.
This values is set as the intensity error of the prominence, i.e. $\hat{\sigma}_{prom}=0.028$.
It should be noted that this value is approximately the same as the standard deviation of the fluctuations shown in the prominence in Figure \ref{intfit} for heights greater than $4$.

The fitting error is determined as the standard deviation of the difference between the normalised intensity and the fit at each pixel.
These errors are shown in Table \ref{gam_table}.
The error bars equivalent to $2\hat{\sigma}_{prom}$ and $2\hat{\sigma}_{fit}$ are shown in Figure \ref{intfit}.
The errors for ${M_*}^2$ from the fitting routine are also shown in Table \ref{gam_table}.

Using the value of $M_*$ from the fitting of the intensity and the definition of $M_*$ as:
\begin{equation}
M_*=\sqrt{\frac{1}{D}}M_{\infty}=\sqrt{\frac{\gamma \beta_{\infty}}{\gamma \beta_{\infty}+2(\gamma-1)}}M_{\infty}
\end{equation}
It is possible to solve for the plasma $\beta$ of the prominence, where $\beta$ is given by:
\begin{equation}
\beta=\frac{2(\gamma-1)}{\gamma}\frac{M_*^2}{M_{\infty}^2-M_*^2}
\end{equation}
The results for the plasma $\beta$ for different values of $\gamma$ are given in Table \ref{gam_table}.
The range of values covers two orders of magnitude from $\beta=0.044 \pm 0.078$ for $\gamma=1.05$ to $\beta=2.666 \pm 0.080$ for $\gamma=2.0$.
However, the value for $\gamma$ in a prominence is likely to be in the range $\gamma=1.4$\,-\,$1.7$, giving a range of plasma $\beta$ of $\beta=0.47 \pm 0.079$ to $1.13\pm 0.080$.
This is a range of only a factor of $\sim2.5$.

\begin{table}
\begin{center}
\caption{Results for $M_*^2$ and plasma $\beta$ (with errors) for various values of $\gamma$}
\begin{tabular}{ c  c  c  c  c c }
\hline
$\gamma$  & $\chi^2$ of fit & $M_*^2$ & $\sigma(M_*^2)$ & $\beta$& $\sigma(\beta)$ \\ \hline
$1.05$     & $1.410$        & $0.61$  & $0.027$         & $0.04$ & $0.078$         \\
$1.1$      & $1.407$        & $0.62$  & $0.027$         & $0.09$ & $0.079$         \\
$1.15$     & $1.404$        & $0.63$  & $0.027$         & $0.14$ & $0.079$         \\
$1.2$      & $1.401$        & $0.63$  & $0.028$         & $0.20$ & $0.079$         \\
$1.25$     & $1.400$        & $0.64$  & $0.028$         & $0.26$ & $0.079$         \\
$1.3$      & $1.394$        & $0.65$  & $0.029$         & $0.32$ & $0.079$         \\
$1.35$     & $1.390$        & $0.65$  & $0.029$         & $0.39$ & $0.079$         \\
$1.4$      & $1.386$        & $0.66$  & $0.030$         & $0.47$ & $0.079$         \\
$1.45$     & $1.382$        & $0.67$  & $0.030$         & $0.56$ & $0.080$         \\
$1.5$      & $1.377$        & $0.67$  & $0.030$         & $0.64$ & $0.080$         \\
$1.55$     & $1.372$        & $0.68$  & $0.036$         & $0.74$ & $0.080$         \\
$1.6$      & $1.367$        & $0.68$  & $0.031$         & $0.86$ & $0.090$         \\
$1.65$     & $1.362$        & $0.69$  & $0.031$         & $0.99$ & $0.080$         \\
$1.7$      & $1.357$        & $0.70$  & $0.031$         & $1.13$ & $0.080$         \\
$1.75$     & $1.351$        & $0.71$  & $0.032$         & $1.30$ & $0.080$         \\
$1.8$      & $1.345$        & $0.71$  & $0.032$         & $1.49$ & $0.080$         \\
$1.85$     & $1.339$        & $0.72$  & $0.032$         & $1.72$ & $0.080$         \\
$1.9$      & $1.333$        & $0.73$  & $0.033$         & $1.98$ & $0.080$         \\
$1.95$     & $1.327$        & $0.74$  & $0.033$         & $2.29$ & $0.080$         \\
$2.0$      & $1.321$        & $0.75$  & $0.033$         & $2.67$ & $0.080$         \\

\hline 
\end{tabular}
 
\label{gam_table}
\end{center}
\end{table}

\section{Discussion}

In this paper, using Equation \ref{dendist} we found that for the range $\gamma=1.4$\,-\,$1.7$ the observed prominence has a plasma $\beta$ of $\beta=0.47 \pm 0.079$ to $1.13\pm 0.080$.
Using Equation \ref{dendist} and the analysis presented in this paper, it is possible to make some predictions on the nature of the density enhancement at the head of a plume:
\begin{itemize}
\item The larger the value of $M_*$ (i.e. the larger $\beta$ and $M_{\infty}$ are), the larger the increase in density at the head of the plume. This is a simple consequence of the greater perturbation from the incompressible state the system receives.

\item The larger the rising plume, the greater the thickness of the region at the head of the plume over which the density increases. This has important implications relating to which plumes this mechanism can be accurately applied. 
\end{itemize}

It is clear from Table \ref{gam_table} that $\gamma$ presents the biggest uncertainty relating to the value of the plasma $\beta$ of the prominence.
It may be expected that by looking at Equation \ref{dendist} the fit may provide some constraints on the value of $\gamma$, but the $\chi^2$ of the fit is almost the same for each value of $\gamma$ therefore statistically it is not possible to do this.
Separate methods that can constrain the value of $\gamma$ would work well in conjunction with this method to determine the plasma $\beta$ of a prominence.

The compressions that are under investigation here happen on timescales of approximately $1$\,Mm\,/\,$10$\,km\,s$^{-1}\sim 100$\,s, which is shorter than the radiative cooling timescales of the prominence ($\sim 1000$\,s), so the compression should be well modelled as adiabatic.
Currently there is no model to describe accurately how the intensity of prominences changes under adiabatic compression.
With this in mind, we will only use the column density, and ignore the effect of heating.
We can at least be confident that the heating will not move the prominence plasma out of the observable temperature range, as the temperature increase is given as $T'=\rho'^{\gamma -1}\sim \rho'^{2/3}$ when $\gamma=5/3$.
Therefore, an increase of the density of $\sim1.4$ (which is approximately what we find in this paper), would increase the temperature from $8000$\,K to $10^4$\,K which is still in the emission range for Ca II H $\sim 5000$\,-\,$10^4$\,K so would not move the emission out of the passband.
A model of the intensity change due to the increase in temperature under these conditions would increase the accuracy of the calculation from plasma $\beta$.

The value for plasma $\beta$ found here has a strong dependence on the hydrodynamic Mach number as the values found were not small enough to make it unimportant.
Therefore, if downflows of prominence material interact with the head of the plume, they would increase the $M_{\infty}$, which would reduce the value of the plasma $\beta$.
In section \ref{MACH} we assume a static prominence, but if we use a prominence background where the material is falling at sound-speed \citep[based on the downflows observed by][]{Chae2010}, we can calculate a new value for the plasma $\beta$ of $\beta\sim0.14$ for $\gamma=1.65$.
Figure \ref{timedist} does not show any obvious downflows interacting with the head of the plume, so the assumption of $\sim11$\,km\,s$^{-1}$ downflows is probably a huge over estimate, but this calculation is shown to highlight the potential importance of downflows when using this method to calculate the plasma $\beta$.
Therefore, when downflows are not included in a calculation, as is the case in this paper, then the plasma $\beta$ value should always be viewed as an upper limit.

Another point that should be noted is the assumption of an irrotational flow as used in Equation \ref{irrot}.
The most important point is that the plume is rising through the prominence material, pushing it out of the way as it rises.
In a compressible medium with finite Reynolds number, the downstream flow (flow behind the plume head) would develop turbulence and a wake, breaking the irrotational assumption.
However, the compression that is being modelled is at the head of the plume, where the flows are much less complex.
There are also observations that suggest prominences are weakly turbulent \citep{LEO2012}, which would imply that to a some extent the irrotational assumption is also broken upstream of the plume.
But it can be expected that to first order the properties of the mean flow are not affected by the turbulence, but the turbulence would be modified by the mean flow potentially changing the characteristics of the turbulence downstream of the plume head, i.e. turbulence distortion.
To summarise, the assumption used is broken downstream, but holds to first order in the area we are interested in.
Therefore, even though we couldn't use this to describe all the flows associated with the prominence plumes, it does apply for the area that is of interest allowing the plasma $\beta$ to be determined.

One interesting point to note is that we have equations that define the velocity around the plume head, as analysed in Figure \ref{plumeheadvel}.
Therefore, analysis of the velocities at the observed plume head (either through Doppler shift analysis or the cork-tracking used in \citet{BERG2010} and \citet{Chae2010} will allow the projection of the plume on the plane-of-sky to be determined.
This would be important as it would remove some uncertainty in the width of the plume head, making the parameters of the fitted ellipse more accurate.
With this information, it should be possible to determine the prominence's plasma $\beta$ with greater confidence and also, through application of knowledge of the 3D magnetic Rayleigh-Taylor instability, estimate the direction of the magnetic field in the prominence.
We plan to present this work in a future paper.

It should be noted that \cite{CAR1996} studied a similar geometry through numerical simulations to investigate the movement of a magnetic cloud through a magnetised atmosphere in relation to CME propagation.
Therefore, the method presented in this paper may be applicable to more MHD phenomena than just prominences, an investigation into what phenomenon this can be applied to may open up some very rewarding research areas.

\bigskip

Hinode is a Japanese mission developed and launched by ISAS/JAXA, with NAOJ as domestic partner and NASA and STFC (UK) as international partners. It is operated by these agencies in co-operation with ESA and NSC (Norway).
The authors would like to thank the staff and students of Kwasan and Hida observatories for their support and comments.
Special mention goes to Drs. H. Watanabe and H. Isobe, as well as Mr. K. Hiroi of Dept. Astronomy, Kyoto University, Dr. M. Cheung and Dr. S. Gunar.
We would also like to thank the anonymous referee, whose comments helped to greatly improve the clarity of the manuscript.
This work was supported in part by the Grant-in-Aid for the Global COE program ``The Next Generation of Physics, Spun from Universality and Emergence'' from the Ministry of Education, Culture, Sports, Science and Technology (MEXT) of Japan.
Part of this work was performed by AH during a visit to the Inter-University Centre for Astronomy and Astrophysics.

\end{document}